\documentstyle[preprint,aps]{revtex}
\begin{document}
\draft
\tighten
\title{HIGHER DIMENSIONAL TAUB-BOLT SOLUTIONS \\ AND \\
THE ENTROPY OF NON COMPACT MANIFOLDS}

\author{Marika Taylor-Robinson \footnote{E-mail:
    mmt14@damtp.cam.ac.uk}}
\address{Department of Applied Mathematics and Theoretical Physics,
\\University of Cambridge, Silver St., Cambridge. CB3 9EW}
\date{\today}
\maketitle

\begin{abstract}
{We discuss the action of a circle isometry group on  
non compact Euclidean Einstein manifolds. We discuss approaches to
a decomposition of the action and entropy for non compact manifolds in terms 
of the characteristics of the orbit space of 
a suitable isometry. There is entropy 
associated with non trivial cohomology of the orbit space of the
isometry, and we consider a class of non compact solutions for which 
such contributions do not vanish. To obtain suitable solutions we 
generalise the Bais-Batenburg construction of
higher dimensional Taub-Nut type solutions to obtain the corresponding 
bolt solutions. We consider the generalisations to non compact
solutions of gravity coupled to scalar and gauge fields.}
\end{abstract}
\pacs{PACS numbers: 04.50.+h, 04.65.+e, 04.70.Dy}
\narrowtext

\section{Introduction}
\noindent

In a previous paper \cite{MMT} we considered a classification scheme
of $d$ dimensional Euclidean Einstein compact manifolds
based on the existence of a one parameter isometry group, in terms of
characteristics of the orbit space of the isometry action.
This work generalised the classification scheme of  
four dimensional manifolds presented in \cite{GH} and 
extended the discussions on fixed point
sets of isometries of \cite{GPP}, \cite{Do2} and \cite{Ra}. 

The main object of our classification scheme was to
extend the geometric interpretation of the entropy in terms of fixed point
sets to general dimensions. What we found was that
$(d-2)$ dimensional bolts have an intrinsic entropy related to their
volume. There are additional contributions to the entropy associated
with non trivial cohomology of the orbit space; when the $(d-3)$
cohomology of this space is trivial we may interpret the entropy
contributions as generalised nut contributions from each fixed point
set. 

Having discussed the action of isometries on Euclidean Einstein
manifolds, it was natural to consider the extensions to Euclidean
solutions of Einstein gravity coupled to scalar and gauge fields. We
found that the same decomposition of the action held, but that if the 
``electric'' part of the gauge field was non vanishing, there would be an
additional term in the action dependent on this part of the field. 
In this context, ``electric'' means that if we consider the action 
of an isometry $\partial_{\tau}$ the 
${\mathcal H}_{\tau...}$ components of the gauge 
field are non zero. If we analytically continue the solution and
$\tau$ is interpreted as an imaginary time coordinate, this part of
the gauge field will indeed be electric. 

\bigskip 

In this paper,
we discuss non compact Euclidean Einstein manifolds admitting 
at least a circle isometry group. The treatment of non compact
solutions in this way is much more subtle for several
reasons. Firstly, one must define all thermodynamic quantities with
reference to a background, and, secondly, unlike compact solutions,
one has to identify an appropriate temperature before one can define
the entropy. That is, we must work within a canonical (or grand
canonical) ensemble. What we find is that although we can find a
partial decomposition of the characteristics of the isometry action,
there is in general a surface term on the boundary term at infinity
which is left over. This surface term will be related to
the energy and angular momentum of the Lorentzian solution with respect 
to the appropriate background. 

For compact manifolds, there is no preferred isometry with respect to
which we should decompose the action. Using different circle subgroups
will give different fixed point set contributions, but the total action
will be the same whichever circle subgroup we consider. Of course, if
we consider a Lorentzian continuation, it may be more
natural, and more physically meaningful, to consider the action
of a time Killing vector. 

For non compact manifolds, although we can define a decomposition of
the action in terms of the characteristics of any circle subgroup,
the boundary terms at infinity will only have a natural physical
interpretation as the energy and angular momentum if one uses an
isometry which has null fixed point sets in the Lorentzian
continuation. We then postulate that the action takes the form
\begin{equation}
S_{E} = \beta E - \beta \omega_{i}J_{i} - S_{f},
\end{equation}
where $\beta$ is the periodicity of the generator of the horizons, $E$
is the energy and $\omega_{i}$, $J_{i}$ are the angular velocity and
momentum. Within a canonical ensemble, we would then interpret the
quantity $S_{f}$ as the entropy; in higher dimensions 
it can be expressed in a similar way to that for compact manifolds \cite{MMT}, 
except that one must consider the action of a suitable Killing vector.
For black $p$-branes, the entropy is 
related to the volumes of the horizons but in general there will be
other contributions arising from non-trivial cohomology of the orbit space. 

\bigskip

We do not attempt to show that the surface terms at infinity can be
related to the mass and angular momentum for a general
solution. Instead we consider a class of static solutions having
a single fixed point set on the Euclidean section which exhibits non
trivial nut behaviour. In four dimensions, the solution that we will
consider is the Taub-Bolt solution \cite{DNP}, 
within a background Taub-Nut solution. 
In higher dimensions, the
analogues of the Taub-Nut solutions had been constructed previously
(the Bais-Battenburg monopole solutions \cite{BB}) and we discuss here the 
construction of the corresponding analogues to the Taub-Bolt
solution. Given such solutions and backgrounds, it is the straightforward to
relate the surface terms to the energy, and explicitly demonstrate that
the action can be decomposed as above. 

Thus we find that these non compact solutions admitting Dirac strings have
an entropy with respect to an appropriate background which is related
not only to the $(d-2)$ volumes of fixed point sets, but also to the
nut behaviour of the fixed point set. Since one cannot have such
contributions for asymptotically flat spacetimes, one might question
the physical relevance of this result. However such spacetimes appear as
exact backgrounds in string theory and should perhaps not be neglected. One
can in addition argue that these solutions are asymptotically flat
in the sense that all components of the Riemann tensor fall off
sufficiently quickly at infinity. Such solutions could also be pair
created, so that there is no net Dirac string singularity. 

\bigskip

It is natural to consider the extensions to solutions of Einstein
gravity coupled to appropriate gauge and scalar fields. Again we find
that the decomposition of the volume term is unaffected, except when
there is an electric gauge field when we will obtain an additional
integral left over. For black brane solutions, one can relate this
term to the electric charge and gauge potential at infinity; the
physical interpretation is that one considers only variations of the
gauge field which leave the electric charge unchanged. 

However, for more general solutions, one cannot necessarily 
define an electric (or magnetic) charge; there may be no topologically
non trivial surfaces over which we can integrate the appropriate
forms. The additional term in the action depends on the electric part
of the field, and is well defined, but cannot be related to a surface
integral. Nevertheless the entropy for such solutions can be defined if
we interpret this term as a constraint. 

As an example of such behaviour we could consider the Israel-Wilson
metrics of \cite{IsWi} but 
instead we consider Euclidean sections of four dimensional heterotic
string theory which are obtained from the Taub-Nut solutions by
T-duality transformations \cite{JMy}. Such solutions are of interest
since they are again exact backgrounds in string theory, obtained by
applying appropriate duality transformations to the Taub-Nut solutions. 
We show that the entropy can be defined,
although the integrals over the gauge fields cannot be expressed in a
simple form. 

\bigskip 

The plan of this paper is as follows. In \S\ref{revie} we review the main 
results of our previous paper. In \S\ref{mkt} we discuss the
action of non-compact solutions and the choice of background.
In \S\ref{bbse} we construct higher dimensional analogues of the four
dimensional Taub-Bolt solution, and in \S\ref{agbs} we calculate 
the action with respect to an appropriate background. We discuss the
analytic continuation of the Bais-Batenburg solutions in \S\ref{lcbb}.

In \S\ref{thnc}, we consider the thermodynamics of non-compact
solutions, and in \S\ref{cgd} we generalise the discussion to
solutions of gravity coupled to scalar and gauge fields. In
\S\ref{dtn} we discuss the application of these arguments to dyonic
Taub-Nut solutions, and we give our conclusions in \S\ref{fin}.

\section{Properties of symmetries} \label{revie}
\noindent

We will consider solutions of the Euclidean action of
$d$ dimensional Einstein gravity
\begin{equation}
S_{E} = - \frac{1}{16\pi G_{d}} \int_{M} d^dx \sqrt{\hat{g}} [R_{d} - m] - 
 \frac{1}{16\pi G_{d}} \int_{\partial M} d^{d-1}x \sqrt{b} 
({\mathcal K} - {\mathcal K}_{0}),
\label{daction}
\end{equation}
where $\hat{g}$ is the determinant of the $d$ dimensional metric
and $R_{d}$ is the Ricci scalar. The $d$ dimensional oriented manifold $M$
will in general have a $(d-1)$ dimensional boundary at infinity 
which we denote as $\partial M$. $m$ is related to the cosmological constant 
$\Lambda$ as $m = (d-2)\Lambda$, and $G_{d}$ is the $d$-dimensional Newton 
constant. ${\mathcal K}$ is the extrinsic curvature of the boundary, and 
${\mathcal K}_{0}$ is the curvature of the boundary of the background with 
respect to which we must measure thermodynamic quantities of non compact 
solutions. 

Many solutions of interest admit continuous symmetry groups of at
least two parameters, and we assume here the existence of at least 
a one parameter group.
A solution admitting a Killing vector $k$ with closed orbits can be
written in terms of $(d-1)$ dimensional fields, which we refer to as
the dilaton $\phi$, gauge potential $A_{i}$ and metric $g_{ij}$, as
\begin{equation}
{ds^2 = e^{\frac{-4\phi}{\sqrt{d-2}}}(dx^d + A_{i}dx^i)^2 +
e^{\frac{4\phi}{(d-3)\sqrt{d-2}}}g_{ij}dx^idx^j},
\label{line-el}
\end{equation}
where we take the Killing vector to be $\partial_{x^d}$ of period
$\beta = 2 \pi \mu$. 
The volume term in the 
action can be expressed in terms of the lower dimensional fields as 
\begin{equation}
S_{E} = - \frac{1}{16\pi G_{d-1}} \int_{\Sigma} d^{d-1}x \sqrt{g} \left[ R -
m e^{\frac{4\phi}{\sqrt{d-2}(d-3)}} - \frac{4}{d-3}(\partial\phi)^2 - 
\frac{1}{4}e^{\frac{-4\sqrt{d-2}}{d-3} \phi}F^2 \right],
\label{xxy}
\end{equation}
where $G_{d} = \beta G_{d-1}$. We refer to 
the $(d-1)$ dimensional manifold we obtain after dividing out by the 
$U(1)$ isometry as $\Sigma$ with $(d-2)$ dimensional boundary
$\partial \Sigma$. Even if the original $d$ dimensional manifold has no 
boundary, the $(d-1)$-dimensional boundary obtained by dividing out by the 
circle action will have boundaries at the fixed points of the circle action;
the total boundary consists of the set of boundaries around each fixed point 
set plus the dimensional reduction of the original boundary. 

\bigskip

If the isometry generated by the Killing vector has fixed point sets,
then the metric $g_{ij}$ will be singular at these points. Denote by
$\mu_{\tau} : M \rightarrow M$ the action of the group where $\tau$ is
the group parameter. At a fixed point, the action of $\mu_{\tau}$ on
the manifold $M$ gives rise to an isometry $\mu_{\tau}^{*} : T_{p}(M)
\rightarrow T_{p}(M)$ where $\mu_{\tau}^{*}$ is generated by the
antisymmetric tensor $k_{M;N}$. Vectors in the kernel $V$ of $k_{M;N}$
leave directions in the tangent space at a fixed point invariant under
the action of the symmetry. The image of the invariant subspace of
$T_{p}(M)$ under the exponential map will not be moved by
$\mu_{\tau}$, and so will constitute a submanifold of fixed points of
dimension $p$ where $p$ is the dimension of the kernel of $V$. Since
the rank of an antisymmetric matrix must be even, the dimension of the
invariant subspace may take the values $0,2,..,d$ for $d$ even, and
$0,.,(d-1)$ for $d$ odd. If the fixed point set is decomposed into
connected components, each connected component is a closed 
totally geodesic submanifold of even codimension \cite{Be}.

\bigskip

We briefly mention here examples of complete non-singular Einstein
manifolds which are of interest physically. In \cite{MMT} we considered 
various homogenous compact Einstein manifolds of positive curvature. 
The only 
non-compact Einstein manifolds that we will consider here are those which
are Ricci-flat; solutions of negative curvature were recently
discussed in \cite{GJ}. 
The simplest examples of non-compact complete metrics admitting at
least a one parameter isometry group are the generalised
Kerr-Newman solutions, constructed in \cite{Per},
characterised by the mass and $[(d-1)/2]$ rotation parameters
(where $[x]$ denotes the integer part of $x$). The general Euclidean
solution has an isometry group $U(1) \times SO(2)^{[\frac{d-1}{2}]}$,
which is enhanced to $U(1) \times SO(d-1)$ in the Schwarzschild limit.
Although the Lorentzian interpretation of these solutions is usually
taken to be rotating black holes in general dimensions, an
interpretation in terms of nucleation of magnetic 
$p$-branes has also been discussed recently \cite{Do2}. 

Asymptotically flat solutions are the most physically relevant
solutions, but more general non-compact vacuum solutions are known. In
four dimensions examples include the Taub-Nut and Taub-Bolt solutions,
and we will consider higher dimensional generalisations in \S\ref{bbse} and 
\S\ref{agbs}. 
In later sections of the paper, we will also consider 
solutions of Einstein gravity coupled to scalar and gauge fields. Once
again we will mostly be interested in solutions of the black hole or
generalised Taub-Nut types.

\bigskip

In our previous paper we rewrote the action in terms of the lower
dimensional fields and an effective potential which we defined. 
For compact manifolds without boundary we were then able to
obtain an expression for the action entirely in terms of the
properties of the fixed point sets. That is, we expressed the volume term 
in the action as 
\begin{equation}
S_{E} = - \sum_{a} \frac{V_{a}}{4 G_{d}} - \frac{\beta}{16 \pi G_{d}}
\int_{\Sigma} F \wedge \bar{G}, \label{tyy}
\end{equation}
where $V_{a}$ is the $(d-2)$ dimensional volume of the $a$th fixed point set, 
and $\bar{G}$ is a $(d-3)$ form which is related to the dual $G$ of 
Kaluza-Klein gauge field $F=dA$ in the $(d-1)$ dimensional metric, as
$G = f \bar{G}$ where 
\begin{equation}
f = \exp(\frac{4\phi\sqrt{d-2}}{(d-3)}).  \label{func}
\end{equation}
For a generic manifold $M$ the form of the cohomology contributions can be 
quite complex but one obtains simplifications when $M$ is a non trivial or 
trivial radial extension 
of a $U(1)$ bundle over a base manifold. The latter class of manifolds
includes complex projective spaces and solutions related to 
cosmological black hole pair creation. 

Now for compact manifolds the entropy is minus the action \cite{GH}, and
thus (\ref{tyy}) gives the decomposition of the entropy in terms of
the boundary volume and cohomology of the orbit space. 
In the context of the no boundary proposal, we can use this
decomposition to define the entropy, and hence the probability of a 
process occurring, in terms of the isometries of the
original Euclidean manifold. 

\bigskip

We extended these ideas to compact solutions of theories involving not 
only the graviton, but also other
fields. We considered a generic action of the form 
\begin{equation}
S_{E} = - \frac{1}{16\pi G_{d}} \int_{M} d^d x \sqrt{\hat{g}} [R 
- e^{-b \Phi} m - (\partial\Phi)^2 - e^{-a\Phi}{\mathcal H}_{p+1}^2],
\end{equation}
where $\Phi$ is the dilaton, and ${\mathcal H}_{p+1}$ is a $(p+1)$ form. 
Depending on the values of $a$, $b$ and $p$, this gives the appropriate action 
for Einstein-Maxwell theories coupled to a dilaton, and for particular limits 
of supergravity theories. 
Upon dimensional reduction, we obtain a $(d-1)$
dimensional $(p+1)$ form ${\mathcal H}_{m}$ and a $(d-1)$ dimensional 
$p$ form ${\mathcal H}_{e}$. We call the former the ``magnetic'' part
of the field, and the latter the ``electric'' part of the field. The
reason for this terminology is that if one
analytically continues the solution, and interprets the Killing
direction as the imaginary time, the resulting gauge fields are electric and 
magnetic. 

In the context of the no boundary proposal, the $(d-1)$
dimensional gauge field arising from the metric must vanish if a
Lorentzian evolution is to exist, since otherwise the Lorentzian
and Euclidean metrics could not both be real. This then implies that
the imaginary time Killing vector has only $(d-2)$ dimensional fixed
point sets, which we interpret as horizons in the Lorentzian 
continuation. 

For pure magnetic fields, we found that the action could be decomposed 
in terms of only the fixed point sets of the imaginary time Killing vector, 
but that for pure electric fields there was an additional volume term 
\begin{equation}
S_{E} =  - \sum_{a} \frac{V_{a}}{4G_{d}} + \frac{1}{8\pi G_{d}} \int_{M}
  d^dx \sqrt{\hat{g}} e^{-a\Phi} {\cal H}_{p+1}^2.
\end{equation}
That is, the action for the solution depends not only on the fixed
point sets, but also on a volume integral of the gauge field. In the 
cosmological context, one adds another term on the initial value hypersurface
which exactly cancels this volume integral; the interpretation is that one
includes only solutions of the same charge in the thermodynamical ensemble.

We will now consider approaches to decomposing the action and entropy of
non compact solutions of both Einstein gravity and gravity coupled to other
fields in terms of the action of an appropriate isometry.  

\section{Non-compact solutions} \label{mkt}
\noindent

Since many interesting Einstein manifolds are non-compact, such as
Euclidean black hole and monopole solutions, it would be useful if
there existed a similar decomposition of the action in terms of the
orbit space characteristics for non-compact solutions. One would not however
expect the action to be expressed solely in terms of these properties of
this orbit space, since this would imply that the action vanishes when the
circle action is trivial which is not necessarily the case for $d > 4$. 
For the compact case we have excluded the possibility of flat circle factors,
and thence the action can be decomposed solely in terms of the orbit
space. 

\bigskip

One can in fact obtain a decomposition of the Euclidean action,
at least in part, using the action of an isometry. As usual we
consider a $d$ dimensional manifold $M$ with a boundary $\partial M$
which admits at least a circle isometry group. Dimensional reduction
along closed orbits of the Killing vector then gives a $(d-1)$
dimensional manifold $\Sigma$ whose boundary can be decomposed into
the boundaries of the fixed point sets $\partial \Sigma_{f}$ and the
dimensional reduction of the original boundary at infinity
$\partial \Sigma_{\infty}$. Now the total volume term in the action is
\begin{eqnarray}
S_{E}^{vol} &=& - \frac{1}{16 \pi G_{d}} \int_{M} d^dx \sqrt{\hat{g}} R; \\ 
&=& - \frac{\beta}{8 \pi G_{d}} \int_{\Sigma} H_{D} = 0  
\end{eqnarray}
where we use the decomposition of the volume term given in \cite{MMT}.
The $(d-1)$ form is defined by 
\begin{equation}
H_{D} = \frac{2}{\sqrt{d-2}} d(\ast d \phi) + \frac{1}{2} F \wedge
\bar{G}.
\label{potgd}
\end{equation}
Let us firstly assume that the $(d-3)$ cohomology of $\Sigma$ is trivial; then
$\bar{G}$ is globally exact, and we can convert the volume integral
into an integral over the boundaries of $\Sigma$ 
\begin{equation}
S_{E}^{vol} = - \frac{\beta}{8 \pi G_{d}} \lbrace
\int_{\partial\Sigma_{f}} J_{D} + \int_{\partial\Sigma_{\infty}} J_{D}
\rbrace = 0,
\end{equation}
where the dilation current is defined by $H_{D} = dJ_{D}$. 
We have imposed the condition that the
solution satisfies the field equations and so the fixed point set and
boundary terms are equal and opposite. It is convenient to define the 
directions of the normals so that 
\begin{equation}
S_{E}^{vol} = \sum_{a} \frac{V_{a}}{4 G_{d}} + \frac{\beta}{16 \pi G_{d}}
\sum_{a}\int_{M_{a}^{d-2}} F \wedge \Psi 
- \frac{\beta}{8 \pi G_{d}} \int_{\partial\Sigma_{\infty}} 
J^{i}_{D} d\sigma_{i}.
\end{equation}
That is, we here define normal vectors pointing out of the manifold to be 
positive, and those pointing into the manifold to be negative. 
Then the total action can be written as 
\begin{equation}
S_{E} = \sum_{a} \frac{V_{a}}{4 G_{d}} + \frac{\beta}{16 \pi G_{d}}
\sum_{a}\int_{M_{a}^{d-2}} F \wedge \Psi + S_{boundary},
\end{equation}
where the new boundary term is given by
\begin{equation}
S_{boundary} = - \frac{\beta}{8 \pi G_{d}} \int_{\partial
  \Sigma_{\infty}} J^{i}_{D} d\sigma_{i} 
- \frac{1}{8 \pi G_{d}} \int_{\partial M} d^{d-1}x 
\sqrt{b} ({\cal{K}} - {\cal{K}}_{0}), \label{bouy}
\end{equation}
and we have added in the surface geometry term.
Let us now consider the question of the background. Since the total volume
term vanishes, we usually match the solutions on the boundary at
infinity, and subtract the background boundary term from that of the
solution in which we are interested. Here however when we 
decompose the volume term we should do the same for the background;
that is, we should express the volume term for the background in terms
of the fixed point sets in the background and an integral of the
background dilation current over the boundary. 

For most physical
solutions, such as black holes, one takes the background to be flat
space, for which the dilation current, and hence the volume
term decomposition, is trivial. For more general solutions, such as
those of the Taub-Nut type, 
the background will not be flat, and we may need to
subtract fixed point set contributions from the background.

One might be concerned that in many solutions the dilaton field
approaches a constant value at infinity, and that a constraint should
be imposed to ensure that this is so. However, it is the choice of
background and the matching of the geometries on the boundary that
will impose this constraint. 

\bigskip

In the general case, when the periods of $\bar{G}$ are non-zero, one
cannot express the action simply in terms of boundary
contributions. Splitting the $(d-1)$ form into dilaton and gauge field
parts, we can however express the action as
\begin{equation}
S_{E}^{vol} = \sum_{a} \frac{V_{a}}{4 G_{d}} + \frac{\beta}{16 \pi G_{d}} 
\int_{\Sigma} F \wedge \bar{G} - - \frac{\beta}{8 \pi G_{d}} \int_{\partial
  \Sigma_{\infty}}  (\partial^{i} \phi) d \sigma_{i},
\end{equation}
where we must subtract the appropriate background quantities. We
must choose the background such that the $d$ dimensional metrics match to
appropriate order on the boundary at infinity. This implies that the
$(d-1)$ forms $F \wedge \bar{G}$ must match between solution and
background to sufficiently high order at infinity that one can regard
the contribution from this term in the integral as arising from inside
the manifolds. 

\bigskip

In general, the boundary term is not zero and the action cannot
be expressed solely in terms of the volume term. One can see
immediately that this must be so by considering two simple
examples. Firstly, as we said above,
one can take any Ricci-flat Euclidean solution
cross a flat circle direction and reduce along the circle; the action
in general does not vanish, but the action of the isometry is
trivial. Secondly, one could take the $d$ dimensional Schwarzschild
solution; the imaginary time Killing vector has a single $(d-2)$
dimensional fixed point set at the horizon, but the action is not one 
quarter of the volume of the event horizon except in four dimensions.
At a physical level, one would expect the entropy, and not the action, to be 
related to fixed point sets of an appropriate Killing vector.  

In special cases, such as the four dimensional Schwarzschild solution
with the Killing vector being the imaginary time direction, 
the sum over fixed point sets will be equal to the
surface geometry term and we can express the action in terms of only
the fixed point sets. We can make further progress in decomposing the
action in terms of the properties of the orbit space, but before
we do so we will consider a class of solutions which are not
asymptotically flat, the higher dimensional generalisations of the 
four dimensional Taub-Nut and Taub-Bolt solutions. 

\section{Bais-Batenburg solutions} \label{bbse}
\noindent

A class of instanton solutions was constructed in \cite{BB} by radially
extending circle bundles over homogeneous K\"{a}hler manifolds. In the
case of non-trivial bundles, the solutions are only regular at the
origin if the K\"{a}hler manifold $M_{K}$ is a complex projective space, the
simplest example being four dimensional Taub-Nut. The trivial bundles 
give rise to regular Euclidean black hole solutions with the topology of
$R^2 \times M_{K}$. More general examples of inhomogeneous metrics on
complex line bundles were constructed in \cite{PP}, with the higher
dimensional Taub-Nut solutions being particular examples. 

Here we discuss the generalisation of the four-dimensional Taub-Bolt
solution, and its relation to the Bais-Batenburg monopole solution. We
also consider the relationship between the latter and solutions
obtained by taking integral powers of the Hopf bundle over complex
projective spaces. 

\bigskip

The form of the metric for a solution of real dimension $(2n+2)$ is 
\begin{equation}
ds^2 = A(r)^2 dr^2 + B(r)^2 (d\tau + A)^2 + C(r)^2 ds_{2n}^2,
\end{equation}
where if $g_{ij}$ denotes the metric on the base manifold, 
the Ricci curvature is taken to be $R_{ij} = \lambda g_{ij} = 2 g_{ij}$. 
The functions are given by 
\begin{eqnarray}
C(r)^2 &=& 2(r^2 - q^2); \hspace{5mm} A(r)B(r) = 2q; \label{bbq}\\
B(r)^2 &=& \frac{4 q^2 r}{(r^2 - q^2)^n} 
[\int^{r}_{q} \frac{(s^2 - q^2)^{n}}{s^2}ds - \alpha ], \nonumber
\end{eqnarray}
with $\alpha$ an integration constant and the range of the radial
coordinate limited to $r$ greater than $q$. Note that our conventions
differ slightly from those in \cite{BB}; the reasons for our choices
will become clear later. Now it was stated in \cite{BB}
that no regular solutions exist unless this integration constant is
set to zero, since if one looks at the behaviour of the proper length of
the circle direction it blows up as $r \rightarrow q$, implying that
the curvature diverges here. 

However,
provided one takes the integration constant to be positive, then there will
exist an $r_{0} > q$ at which the proper length of the circle
direction degenerates to zero. That is, there exists a solution 
$r_{0} > q$ to 
\begin{equation}
\int^{r_{0}}_{q} \frac{(s^2 - q^2)^n}{s^2} ds - \alpha = 0,
\end{equation}
and we can define the radial coordinate to extend from $r_{0}$ to
$\infty$. The bundle structure breaks down at $r_{0}$ in that the
radius of the circle goes to zero, and seals off the boundary; it is
this which ensures that the manifold is regular and geodesically complete.

This follows if we express the integral as
\begin{equation}
\int^{r}_{q} \frac{(s^2 - q^2)^n}{s^2} ds = \frac{1}{r} p_{2n}(r),
\end{equation}
where $p_{2n}(r)$ denotes a polynomial of order $2n$, of which the 
only properties we need to know are that
\begin{equation}
p_{2n}(q) = 0, \hspace{3mm} p_{2n}(r>q) > 0.
\end{equation}
Thus, for any positive $\alpha$, there must exist at least one
solution of $p_{2n}(r_{0}) = \alpha r_{0}$, with $r_{0} > q$. 
Since $p_{2n}(r)$ is also a monotonically increasing function for $r > q$,
there exists precisely one solution $r_{0}$ for each value of $\alpha$.

If we take $\alpha$ to be zero, the Killing vector
$\partial_{\tau}$ will have a nut point singularity at the origin of
the coordinate system $r = q$. If however $\alpha$ is not zero, then 
there will be a bolt of dimension $(d-2)$ at the corresponding origin
of the coordinate system $r = r_{0}$. Hence, it is easy to see that
the latter solutions are likely to be the
higher dimensional analogues of the Taub-Bolt solution in four dimensions.
One can extend the analysis of \cite{BB} to show that the
solutions of general $\alpha$ are regular at the origin of the
coordinate system only if the base manifold is a complex projective space. 

\bigskip

However, non-singular solutions cannot be defined for all positive
$\alpha$; only for a subset of parameters will conical singularities
at the origin be eliminated. The
periodicity of the circle direction is determined by looking at the
behaviour of the metric in the vicinity of the fixed point set.
For the nut solution, in the vicinity of the fixed point, we can bring the
metric into the form
\begin{equation}
ds^2 = d\rho^2 + \frac{\rho^2}{(n+1)^2} d\tau^2,
\end{equation}
where we consider a two-dimensional subspace obtained by 
fixing the coordinates on the projective space. Evidently regularity
at the nut then requires 
\begin{equation}
\beta = 2\pi(n+1).
\end{equation}
Now in \cite{PP} the periodicity of a regular solution of this type
was found to be 
\begin{equation}
\beta = \frac{4\pi p}{k \lambda},
\end{equation}
where $p$ is an integer such that the Chern class of the tangent bundle 
evaluated on $H_{2}(M_{K},Z)$ is $Z \cdot p$, $k$ is an arbitrary integer
and $\lambda$ is the curvature of the base manifold, as defined previously.
For a complex projective space, $p = (n+1)$ and so our answer is in
agreement with this general formula with $k=1$. The nut manifold is
topologically $R^{2n+2}$ with the length of the Hopf circles 
asymptotically tending to a constant, whilst the base expands.

For the bolt solutions, the metric in the vicinity of the fixed point
set $r = r_{0}$ takes the form
\begin{equation}
ds^2 = d\rho^2 + \rho^2 (\frac{q}{r_{0}})^2 d\tau^2,
\end{equation}
where we again consider a two-dimensional subspace. If we suppose that
the periodicity of the circle direction is 
\begin{equation}
\beta = \frac{2\pi (n+1)}{k},
\end{equation}
then conical singularities at the bolt are eliminated provided that
$r_{0} = q(n+1)/k$. In addition, for the solution to be regular we
require that the proper length of the circle direction degenerates to
zero for some $r_{0} > q$ and so it follows that solutions exist for
$k < (n+1)$. In fact, there also exists a non-singular solution in the
limit that $k = (n+1)$, as we might expect. 

\bigskip

The solution for which $k=1$ is in some sense singled out, because the
periodicity of the circle direction is identical to that in the nut
solution. In physical terms this will mean that the Dirac string type
behaviour associated with the bolt is identical to that in the nut
solution; the manifold is again the first
power of the Hopf bundle over the base manifold.
In the limit that $n=1$ we return to the well-known
Taub-Bolt solution first constructed by Page \cite{DNP}. 
Note that neither the generalised Taub-Nut nor the generalised Taub-Bolt
solution are asymptotically locally flat unless $n=1$; we shall return
to this point later. For more general $k$, we obtain solutions which 
are the $k$th power of the Hopf bundle over the fixed point set.

If $k=0$ the solution degenerates to a trivial
bundle over the base manifold which is equivalent to a generalised
Euclidean black hole solution. We can show this by taking the limit 
$k,q \rightarrow 0$, with $r_{0}$ finite. After a little manipulation
we find
\begin{equation}
ds^2 = (1 - (\frac{a}{R})^{2n-1})d\tau^2 + 
\frac{dR^2}{(1 - (\frac{a}{R})^{2n-1})} + R^2 d\bar{s}^2_{2n},
\end{equation}
where we have rescaled the metric on the base manifold so that 
$R_{ij} = (2n-1)g_{ij}$ and the radius at infinity is one. Note that
although in the non-degenerate solution we required the base manifold
to be a complex projective space for regularity we can drop
this condition for the degenerate solutions, since we can certainly
take it to be a sphere. In fact one can show that any compact
Einstein base manifold will give a regular solution \cite{BB}. 

\bigskip

As stated above, there is also an appropriate limit in which we can obtain a
regular metric from the $k = (n+1)$ solution; this was shown for $n=1$
in \cite{DNP}. Firstly, we choose
\begin{equation}
(r_{0}^2 - q^2) = a^2,
\end{equation}
where $a^2$ is an arbitrary positive constant. Now regularity of the
solution requires that $2\pi r_{0}/q = 2\pi$, and thus we must take
both $r_{0}$ and $q$ to infinity. One then defines 
\begin{equation}
R^2 = (2n+2)(r^2 - q^2),
\end{equation}
which implies that as we take the limit that $q \rightarrow \infty$
the metric becomes 
\begin{equation}
ds^2 = \frac{4R^2}{(2n+2)^2} (1 - (\frac{a}{R})^{2n+2}) (d\tau +
A)^2 + \frac{dR^2}{(1 - (\frac{a}{R})^{2n+2})} +
\frac{R^2}{2(n+1)}ds^2. 
\end{equation}
where the periodicity of the circle direction is $2\pi$. Now for $n=1$
this metric is easily recognisable as the Eguchi-Hanson metric. In
fact for $n>1$ the metric coincides with that given by
Calabi \cite{CA} and others \cite{GZ}. In addition, for this
solution, the integers $k$ and $p$ which we defined previously are the
same which is precisely the condition required for the solution to be
K\"{a}hler and to take this simple form \cite{PP}. Note that this
solution is asymptotically locally Euclidean for all $n$. 

\bigskip

Going back to the $k=1$ solution, the generalised Taub-Bolt solution,
the behaviour of the metric close to the bolt is the
same as the behaviour close to the $(d-2)$ dimensional bolt 
in $CP^{n+1}$. One can express the metric for the latter as \cite{MMT}
\begin{equation}
ds^{2} = 2(n+1) \Lambda^{-1} \lbrace d\theta^{2} + \sin^{2}\theta 
\cos^{2}\theta (d\tau - A)^{2} + \sin^{2}\theta d\bar{s}_{2(n-1)}^{2}
\rbrace, \label{clmet}
\end{equation}
with endpoints at $\theta = 0$ and $\theta = \pi/2$. We choose the
normalisation of the metric on $CP^{n-1}$ such that $R_{ij} = 2n
g_{ij}$, and $dA$ can be chosen as the K\"{a}hler form on $CP^{n-1}$. 
Then the resulting metric is isometric to the standard Fubini Study metric on 
$CP^{n}$. In particular, the Killing vector $\partial_{\tau}$ has a 
nut at the ``origin'' $\theta = 0$ and a $CP^{n-1}$ bolt at ``infinity''
$\theta = \pi/2$. 

Now in the neighbourhood of the fixed point set in the bolt solution, the 
metric takes the form
\begin{equation}
ds^2 = d\rho^2 + (\frac{\rho}{n+1})^2 (d\tau + A)^2 + 2n (n+2) q^2
ds_{2n}^2,
\end{equation}
with the scale of the metric on the base manifold being $R_{ij} = 2g_{ij}$
whilst in the neighbourhood of the $(d-2)$ bolt in $CP^{n+1}$ the metric
takes the form 
\begin{equation}
ds^2 = 2(n+2)\Lambda^{-1} \lbrace d\rho^2 + \rho^2 (d\tau + A)^2 +
d\bar{s}^2_{2n} \rbrace,
\end{equation}
with the scale of the metric in the base manifold being $R_{ij} =
2(n+1)g_{ij}$. So after rescaling the metric on the base manifold we
find that the circle directions have the same periodicity. 

However, the isometry in $CP^{n+1}$
also has a nut fixed point at $\theta = 0$, and so the
topology of the Taub-Bolt is that of 
($CP^{n+1} - \lbrace {\rm point} \rbrace)$ 
which we as usual denote as $(CP^{n+1} - \lbrace 0 \rbrace)$. 
Thus the manifold has an Euler number
of $\chi = n+1$. We would of course expect this behaviour as an
extension of the well-known behaviour for $n=1$ \cite{GibPop}.
The solutions for general $k$ have topology $(CP^{n+1} - \lbrace 0
\rbrace)/Z_{k}$  and again have Euler number $\chi = n+1$. 

\section{Action of generalised bolt solutions} \label{agbs}
\noindent

It is interesting to calculate the action for the generalised bolt
solutions. To find the action of the generalised 
Taub-Bolt solution for which $\alpha \neq 0$, we
must match it to an appropriate background; the natural choice 
is the $\alpha = 0$ solution. That is,
we match the solution to a background with an
equivalent Dirac string type singularity, a background
with equivalent magnetic behaviour. 

Now the natural way 
to match the bolt geometry to the nut geometry on an arbitrary
surface is to rescale the nut parameter in the nut solution. That is, 
we let $q \rightarrow m q$ so that the form of the metric becomes
\begin{equation}
ds^2 = B^2_{0}(mq) (d\tau + A)^2 + A^2_{0}(mq) dr^2 + (r^2 - m^2
q^2)ds_{2n}^{2}, \label{gbr}
\end{equation}
where we denote with subscripts the background
quantities evaluated with $\alpha = 0$.
The constant $m$ is defined so that the induced metric 
on a surface of arbitrary large radius $R_{\infty}$ 
matches to sufficient order, sufficient being up to terms of order 
$1/R_{\infty}^{2n-1}$. Terms of higher order need not match since they
do not contribute to the action. The choice of $m$ required is
\begin{equation}
m^2 = (1 - \frac{\alpha(2n-1)}{R_{\infty}^{2n-1}}).
\end{equation}
It is straightforward to show that the proper lengths of the circle at
infinity then match up to to the requisite order by expanding the
polynomials $B^{2}_{0}(mq)$ and $B^2_{\alpha}(q)$. Retaining only
terms up to the requisite order, the former is 
\begin{eqnarray}
B^{2}_{0}(mq) &\simeq & \frac{4q^2}{(R_{\infty}^2-q^2)^{n}} 
(1 - \frac{\alpha(2n-1)}{R_{\infty}^{2n-1}}) p_{2n}(R_{\infty},q)
\nonumber \\
& \simeq & \frac{4q^2}{(R_{\infty}^2-q^2)^{n}} (p_{2n}(R_{\infty},q) - \alpha
R_{\infty}),
\end{eqnarray}
which is equivalent to $B^2_{\alpha}(q)$.

The bolt term in the action is given by 
\begin{eqnarray}
{\cal K} \sqrt{c} &=& \frac{1}{A_{\alpha}}
\frac{\partial}{\partial r} ( C^{2n}_{\alpha} B_{\alpha}
\sqrt{g_{2n}}); \nonumber \\
& \simeq & 2^{n} q \sqrt{g_{2n}} (\gamma - \alpha), \label{dct},
\end{eqnarray}
where in the latter expression we give only the single term which is
independent of the radius at infinity and $\sqrt{g_{2n}}$ is the
volume element on the base manifold. 
The quantity $\gamma$ is defined as the term independent of
$R_{\infty}$ in the expression
\begin{equation}
[(\int^{R_{\infty}}_{q} \frac{(s^2 - q^2)^n}{s^2} ds)].
\end{equation}
The corresponding term in the nut background is given by
\begin{equation}
{\cal K}_{0} \sqrt{c} \simeq 2^{n} q \sqrt{g_{2n}} (\gamma),
\end{equation}
where we again give only the constant term. The forms of the bolt
solution and its background indicate that the divergent terms will
cancel, and it can be verified explicitly that this is the case. 
Then the total action is given by the simple expression
\begin{equation}
S_{E} = \frac{q (n+1)}{4 G_{d}} \alpha \bar{V}_{2n},
\end{equation}
where $\bar{V}_{2n}$ is the volume of the base manifold, in a 
(rescaled) metric such that $R_{ij} = g_{ij}$. 
Evidently the restriction to positive $\alpha$ ensures that the 
action is positive. 

Let us show now that our general calculation does give the correct
answer for the action of the four dimensional Taub-Bolt solution with 
respect to the Taub-Nut solution. In four dimensions, the scalar
function takes the form
\begin{equation}
B(r)^2 = \frac{4 q^2}{r^2 - q^2} [r^2 - 2 qr + q^2 - \alpha r].
\end{equation}
Since for the Bolt solution, the fixed point lies at $r_{0} = 2q$, 
we find $\alpha = q/2$. This is in agreement with the form of the
Taub-Bolt metric given in \cite{DNP} and \cite{GH}. We can
then calculate the action as
\begin{equation}
S_{E} = \frac{\pi q^2}{G_{4}},
\end{equation}
which is in agreement with the calculation of \cite{cjh_1}. 
It is interesting to
note that this is a solution for which the action {\it can} be expressed
solely in terms of the fixed point sets. Evaluating the potentials
$\Psi$ explicitly, it is easy to show that 
one can choose them to vanish on the boundary; then the only remaining
parts of the dilation current are the dilaton terms. Using the
expression for this part of the current (\ref{potgd}), one can then show
that the net boundary contribution is zero and the action is given by
the sum over the fixed point sets. 

\bigskip

Since the solutions are Ricci-flat, they 
have an interpretation as generalised monopoles when
we add a flat time direction and consider a Kaluza-Klein type
reduction along the circle isometry. The $n=1$ monopoles are of course
well-known \cite{GPer} and \cite{RDS}.
One might think that the action would in some
sense determine the probability of nucleation of
a generalised ``bolt'' monopole within a background ``nut'' monopole. 
There are however several objections to this interpretation. Firstly,
as was pointed out in \cite{Do2}, one can argue that these
objects should not exist in isolation, since the monopoles cannot be
regarded as circle bundles over flat space asymptotically, except for
the Taub-Nut and Taub-Bolt solutions. 

Secondly, to determine whether generalised Taub-Bolt can decay into generalised
Taub-Nut we turn to cobordism theory. That is, following the methods
of \cite{BGR} we glue together these manifolds at infinity and
ask whether there exists a 
cobordism which preserves the Pontryagin and Stiefel-Whitney
numbers. If it does not, we conclude that 
even though the solutions have the same behaviour at infinity, 
generalised Taub-Bolt cannot decay into generalised Taub-Nut.

Now for $n$ odd we know that the bolt solution does not admit a spin
structure, since $CP^{n+1}$ only admits a spin structure when $n$ is
even (see for example \cite{EGH}). 
So the cobordism is necessarily excluded by the non-preservation
of the second Stiefel-Whitney number. For $n$ even one can show that the 
Pontryagin numbers are not conserved and hence the decay is still excluded.
So, although the generalised bolt monopole has a well-defined 
higher mass than the Bais-Batenburg monopole, it cannot decay into the latter.

Of course the cobordism arguments do not exclude the possibility that the bolt
solutions decay into nut solutions plus a solution with zero magnetic
behaviour and compensating Stiefel-Whitney and Pontryagin numbers. 
If one assumes that neither type of monopole exists in isolation one
cannot exclude the possibility of pair creation of the bolt
type monopoles, such that the net Dirac string type singularity vanishes. 
Pair creation of the Bais-Batenburg monopoles is certainly known \cite{Do2}. 

\bigskip

We mention here that the appropriate background for the generalised
Eguchi-Hanson metric is any other such metric with parameter $a'$;
such a background gives the required magnetic behaviour, and the
parameter is arbitrary. The
action of the solution with respect to the background vanishes; we
would expect this, since if we scale the metric by a factor $b$, the
action scales by some (dimension dependent) positive power of $b$. For
the solutions such that $1 < k < (n+1)$, there appears to be no
appropriate background with respect to which we can evaluate the
action. 

\section{Lorentzian continuation of Bais-Batenburg solutions} \label{lcbb}
\noindent

We have so far considered the Euclidean section of the Bais-Batenburg
solutions.
The usual Lorentzian interpretation is to take the product with a flat time 
direction, but we can also continue the four dimensional Euclidean solutions
to obtain four dimensional 
Lorentzian Taub-Nut solutions. This analytic continuation is also
possible for the higher dimensional solutions. Starting with the Euclidean 
metric, to obtain the Lorentzian solution, one should let the ``nut'' 
and ``mass''parameters become imaginary. Let us firstly analytically continue 
$q \rightarrow iq$. Then
\begin{equation}
B(r)^2 \rightarrow - \frac{4q^2 r}{(r^2 + q^2)^n} \lbrace
\int^{r} \frac{(s^2 + q^2)^n}{s^2} ds - \int^{iq} \frac{(s^2 + q^2)^n}{s^2} ds
- \bar{\alpha} \rbrace,
\end{equation}
where the first integral in the brackets is pure real. The latter two 
terms are pure imaginary, since in the Euclidean solution they depend on 
the nut parameter as $q^{2n-1}$. Since these terms vary with the radius
as $r^{1-2n}$, they determine the mass, which we must also analytically
continue to obtain a real Lorentzian evolution. The resulting solution is
\begin{equation}
ds^2 = B_{l}(r)^2 (d\tau + A)^2 + A_{l}(r)^2 dr^2 + (r^2 + q^2) 
d\bar{s}^2_{2n},
\end{equation}
where $A_{l}^2 B_{l}^2 = - 4 q^2$ and 
\begin{equation}
B_{l}^2 = - \frac{4 q^2 r}{(r^2 + q^2)^n} \lbrace \int^r
\frac{(s^2 + q^2)^n}{s^2} ds - \int^{q} \frac{(s^2 - q^2)^n}{s^2} ds - \alpha
\rbrace. \label{bll}
\end{equation}
The periodicity of the time coordinate is still fixed at $\beta = 2 \pi (n+1)$
to ensure that the Dirac string singularity is removable. In the 
Euclidean solution, regularity at the origin, the fixed point, required that
$\alpha$ could only take two values, zero and one other fixed
non-zero value. We can
however obtain a regular Lorentzian solution with any value of $\alpha$. 

\bigskip

The polynomial $B_{l}^2$ has zeroes at two values of $r$, which 
correspond to horizons. We can demonstrate this as follows. 
The term in brackets in (\ref{bll}) can be expressed in the form
\begin{equation}
\frac{1}{r}(\delta(r) - a q^{2n-1} r - b q^{2n}) \label{ply}
\end{equation}
where $a, b$ are (positive definite)
constants which are determined, and $\delta(r)$ is an
even polynomial of order $2n$ having positive coefficients. Then for all $q$
and $\alpha$ (\ref{ply}) will have two roots, one at positive $r$ and the other
at negative $r$; these define the zeroes of the polynomial.

The roots cannot coincide unless $q=0$ and $a q^{2n-1}$
is finite, which corresponds to
the limiting case of a black hole solution for which the
bundle over the base manifold is trivial. It is straightforward to demonstrate
that one can find suitable coordinates such that the metric is non-singular 
at these points, and that they are null horizons. 
In analogy to the four dimensional
solution, one would expect that the interior region $r_{-} < r < r_{+}$
has an interpretation as a cosmological solution for a universe with the
spatial topology of a $U(1)$ bundle over $CP^n$; this is indeed so.

As in the four dimensional solutions (discussed in \cite{HE}), 
the region $r_{-} < r < r_{+}$ has no closed timelike curves, but
there are for $r < r_{-}$ and for $r > r_{+}$. One family of null
geodesics crosses both horizons $r= r_{-}$ and $r=r_{+}$, but the
other family spirals round near these surfaces and is incomplete. 
The surfaces which are the surfaces of transitivity of the isometry
group are spacelike surfaces in the region $r_{-} < r < r_{+}$ and are
timelike for $r > r_{+}$ and $r < r_{-}$. The two surfaces of
transitivity $r=r_{-}$ and $r = r_{+}$ are null surfaces and they form
Cauchy horizons of any spacelike surface contained in the region
$r_{-} < r < r_{+}$, because there are timelike curves in the regions
$r < r_{-}$ and $r > r_{+}$ which do not cross $r = r_{-}$ and $r =
r_{+}$ respectively. The region of spacetime $r_{-} \le r \le r_{+}$
is compact yet there are timelike and null geodesics which remain
within it and are incomplete. 

\bigskip

Note that when we analytically continue the nut solutions, we find that
the former fixed point sets are now contained within the null surfaces and
are non singular points within the spacetime. This is in contrast to black
hole solutions for which the fixed points in the Euclidean solution correspond
directly to horizons within the Lorentzian solution. One could regard the
event horizon as the blowing up of the fixed point surface in the Euclidean
solution; we will then interpret the entropy of the spacetime as being
contained within this surface. 

\section{Thermodynamics of non-compact solutions} \label{thnc}
\noindent

For compact solutions, we have considered only fixed point sets which are in
some sense at the boundaries of the $d$ dimensional manifold. That is, the
fixed point sets arise as non singular origins of coordinate systems which we
use to cover the manifold. We might ask whether it is possible to have fixed
point sets of an isometry which are not origins of coordinate systems. There
are two objections to this possibility for compact manifolds. Firstly, 
such fixed point sets are likely to be associated with physical singularities,
rather than removable coordinate singularities. Secondly, one would not
expect a submanifold through which one can pass freely back and forth to 
a different part of the manifold to be associated with entropy. 

For non compact manifolds, the implications are slightly more subtle. It is
possible for a fixed point set to be non compact; the obvious example is
an acceleration horizon. Such a fixed point set need not be an origin of a
coordinate system as such, but is still associated with entropy and a
temperature. Although we have not discussed non compact fixed point sets
specifically, the analysis given here still applies. For example, although the
volume of the acceleration 
horizon may be infinite, a suitable background will have 
a corresponding horizon which is also infinite in extent, and the finite
difference between the volumes is the quantity which will be physically 
significant. 

If however one has a compact fixed point set embedded in a non compact
manifold, such a fixed point set can only be directly
associated with entropy if it is a boundary of the non-compact manifold. 
If one can pass freely back and forth across the fixed point set into a 
different part of the manifold, there can be no entropy associated
with this fixed point set. One could consider a Euclidean Kerr solution, with
conventional Lorentzian interpretation as a rotating black hole; 
the fixed point set of the imaginary time Killing vector determines neither
the entropy nor the temperature. The physically significant Killing vector
determining these quantities is the isometry which has a fixed point set at the
inner boundary of the Euclidean manifold. 

\bigskip

In \S\ref{mkt} we gave a decomposition of the Euclidean action if a non compact
solution partially in terms of the action of the isometry. This 
decomposition is valid whatever the choice of Killing vector when the symmetry
group is more than one dimensional. However a particular choice of Killing
vector will be usually be singled out; for the Kerr solution, it is the 
Killing vector that has a zero on the inner boundary. One expects this fixed
point set to be associated with entropy, and it is hence useful to decompose
the action in terms of this isometry. 

Suppose we then consider the Lorentzian continuation of such a solution. 
The periodicity of the isometry that has a zero on the inner boundary 
in general determines the temperature of the Lorentzian solution, provided 
that the choice of Lorentzian continuation is such that the inner boundary is 
null. This will usually require that the generator of the horizon contains the
time Killing vector. Note that we assume that the spacetime admits a 
Killing vector which is timelike at infinity; without such a Killing vector one
cannot have a (precise) definition of energy. If on the other hand we consider
a continuation such that the inner boundary remains spacelike, but a
non-compact fixed point set becomes null, the periodicity of the isometry
generating this fixed point set will determine the temperature. 

Again a good example is a Euclidean Kerr solution. One usually
analytically continues the solution to obtain a Lorentzian rotating 
black hole. Under such a continuation, the inner boundary to the
Euclidean solution becomes a null horizon, and the periodicity of the
isometry which leaves this surface fixed determines the temperature of
the black hole. One can however find another analytic continuation;
the resulting Lorentzian solution has an interpretation as pair
creation of monopoles within a magnetic background \cite{Do2}. In this
continuation the fixed point set which becomes null is an acceleration
horizon whose temperature is again determined by the periodicity of
the isometry on the Euclidean section. 

\bigskip

Once we have chosen the analytic continuation we can decompose the action in
terms of the isometry which admits null fixed point sets in the Lorentzian
evolution. The periodicity of this isometry then defines the temperature as
$\beta = 1/T$. In \S\ref{mkt} we gave one decomposition of the action in 
terms of the fixed point sets of the isometries. The more natural decomposition
is perhaps to reverse the choice of normal directions so that
\begin{equation}
S_{E}^{vol} = - \sum_{a} \frac{V_{a}}{4 G_{d}} - \frac{\beta}{16 \pi G_{d}} 
\int_{\Sigma} F \wedge \bar{G} - \frac{\beta}{16 \pi G_{d}} 
\int_{\partial \Sigma_{\infty}} (\partial \phi^{i}) d\sigma_{i}.
\end{equation}
One might then expect that the surface terms at infinity are related to the
energy and rotation of the solution, and that the total action can be
written as 
\begin{equation}
S_{E} = \beta E - \beta \sum_{i} \omega_{i} J_{i} - \sum_{a} \frac{V_{a}}
{4 G_{d}} - \frac{\beta}{16 \pi G_{d}} \int_{\Sigma} F \wedge \bar{G}, 
\end{equation}
where $E$ is the energy, $J_{i}$ are independent conserved angular momenta and
$\omega_{i}$ are angular velocities. The conserved quantities must be
defined with respect to suitable backgrounds, and we must also
subtract background quantities from the sum over fixed point sets, and
integral over the gauge field. The justification for such an 
expression arises from introducing a grand canonical ensemble, and interpreting
the action as a thermodynamic potential. The entropy would then be
given by
\begin{equation}
S = \sum_{a} \frac{V_{a}}{4G_{d}} + \frac{\beta}{16 \pi G_{d}}  
\int_{\Sigma} F \wedge \bar{G}. 
\end{equation}
When $\bar{G}$ has trivial periods, we can interpret the entropy
solely in terms of the fixed point sets as 
\begin{equation}
S = \sum_{a} \frac{V_{a}}{4G_{d}} + \frac{\beta}{16 \pi G_{d}}  
\sum_{a} \int_{M^{d-2}_{a}} F \wedge \Psi. 
\end{equation}
This expression is of course well known for $(d-2)$ dimensional fixed point
sets, and is the same as for compact solutions \cite{MMT}, 
except that we cannot consider
just any fixed point sets, but rather only those than can be surrounded by a 
(compact) null $(d-2)$ dimensional hypersurface or those that are a 
non-compact null $(d-2)$ dimensional hypersurface. 

One might question whether constraint terms relating to nut charge, or
more generally to non-trivial cohomology of $\Sigma$, should be included in the
path integral. However, like magnetic charge, nut charge is fixed by
the boundary conditions; the path integral runs over metrics with
given boundary conditions but the same nut behaviour. On the other
hand, they can have any mass, and so the sum is weighted by a factor 
$exp( - \beta H)$.   

\bigskip

It is not obvious how one
could verify that the surface terms give the entropy and angular momentum in
general without considering specific types of solutions. Since for a generic
solution and background, the definition of the energy is highly
non-trivial \cite{BKKLS}, one would not expect it to be trivial to relate
the surface integral to the energy. We can
however demonstrate that this expression holds for the generalised
Bais-Batenburg solutions discussed in \S\ref{bbse}, for which the angular
momentum vanishes. We could of course also consider the rotating
generalisations of these solutions; the $n=1$ solutions were
constructed in \cite{GibPer}.  
Consistency with the known expressions for static and rotating
black holes in general dimensions, as we take the nut parameter to
zero and rotate the solution, would then imply the general result. 

\bigskip

One would expect to be able to
define the mass of the bolt solution with respect to that of the nut solution
on a surface of constant time by looking at the behaviour of the $r^{1-2n}$
term in $B_{l}(r)^2$, where we rescale 
$B_{l}(r)^2$ so that the circle at infinity has unit radius. 
The normalisation of the mass is fixed from that of the action; for a
Schwarzschild solution with the $\hat{g}_{tt}$ term in the metric
being $\mu r^{1-2n}$ the mass with respect to a background of $\mu =0$ is
\begin{equation}
M = \frac{2n V_{2n}}{16 \pi G_{d}} \mu,
\end{equation}
where $V_{2n}$ is the volume of the base manifold, usually taken to be
a sphere.

Although one might be concerned about the validity of defining the mass
for a solution with such Dirac strings in this way, one can verify that this
approach in fact gives the same answer that one gets by taking into 
account the non trivial fibration of the time coordinate. 
That is, we find that the mass of the generalised
bolt solution with respect to the generalised nut solution is
\begin{equation}
M = \frac{2n \alpha}{16 \pi G_{d} } \bar{V}_{2n},
\end{equation}
where $\bar{V}_{2n}$ is the volume of the base manifold
in the rescaled frame as defined in \S\ref{agbs}. 
Since the periodicity of the circle coordinate is given by 
$\beta = 4 \pi q (n+1)$ when we choose the radius of the circle at
infinity to be one, there is a relationship between the action and the 
mass
\begin{equation}
S_{E} = \frac{\beta M }{2n}. \label{upsid}
\end{equation}
We would indeed expect this. The action $S_{E}$ is given by 
$k \beta^{2n}$, with $k$ a constant, since dimensional
analysis requires that $\alpha$ is proportional to 
$\beta^{2n-1}$. Then, introducing a canonical ensemble,
the mass is given by
\begin{equation}
M = \frac{\partial S_{E}}{\partial \beta} = 2n k \beta^{2n-1},
\end{equation}
which implies the relation (\ref{upsid}). 

\bigskip 

We will now demonstrate the relationship between the integral of the
dilation current over the boundary at infinity and the mass. The integral 
of the $(F \wedge \Psi)$ term in the dilation current over the
boundary at infinity does not contribute; as one might expect, the
integral taken over a surface $r=R$ 
in the solution with respect to an appropriate
background falls off as $1/R^{p}$,
where $p$ is positive, but the proof is slightly subtle. The
effective $(d-1)$ dimensional Einstein frame metric is 
\begin{equation}
ds_{d-1}^{2} = B(r)^{\frac{2}{d-3}} \lbrace A(r)^{2} dr^2 + (r^2 -q^2)
d\bar{s}_{2n}^{2} \rbrace,
\end{equation}
with the $(d-1)$ dimensional two form being $F = dA$. The $(d-3)$ form
is defined by
\begin{equation}
G = 2q B^{-\frac{2}{d-3}} (r^2 - q^2)^{n-2} (\ast F),
\end{equation}
where we take the dual in the metric on the base
manifold. Using the defining relation for the potential,
\begin{equation}
d\Psi \propto \lbrace \frac{(p_{2n}(r) - \alpha r)}{(r^2 - q^2)^2}
\rbrace (\ast F),
\end{equation}
where we omit constant factors for simplicity. 
Note that we can omit the scaling factor $m^2$ for the background
since it will not contribute to terms of sufficiently high order to be of
interest here. To evaluate the integral on a surface of constant $r$,
we require only the $\Psi_{i_{1}..i_{2n}}$ components of $\Psi$, where
$x_{i_{j}}$ are coordinates on the base manifold. These are found from
\begin{eqnarray}
\Psi_{i_{1}..i_{2n}} & \propto &
\lbrace \int \frac{(p_{2n}(r) - \alpha r)}{(r^2 - q^2)^2} dr
\rbrace (\ast F)_{i_1..i_{2n}}; \label{bic} \\
& \propto & \lbrace a_{2n-3} r^{2n-3} + a_{2n-5} r^{2n-5} +
 ... - \frac{\alpha}{2r^2} + ... \rbrace (\ast F)_{i_1..i_{2n}}, \nonumber
\end{eqnarray}
where $a_{i}$ is the coefficient of the term in $r^{i}$. Note that since the 
polynomial $p_{2n}(r)$ is proportional to $(r-q)^{n+1}$, the integrand
is indeed finite at $r=q$ in the background. 
In both the solution and the background the potential
diverges at infinity unless $n=1$; however, 
the leading order contribution to the integral is 
\begin{equation}
\int_{r=R} F \wedge \Psi_{\alpha} - \int_{r=R} F \wedge \Psi_{0}
\propto (-\frac{\alpha}{2R^2}) \int_{M_{K}} F \wedge (\ast F),
\end{equation}
which vanishes in the limit that $R \rightarrow \infty$. Even though
$F$ and $\ast F$, as the unique two and $2(n-1)$ forms on the base
manifold respectively, have Dirac string type singularities, the
integral is well defined, and is proportional to the volume
of the base manifold. Note that we
have assumed that the integration constant in (\ref{bic}) matches
between solution and background.

\bigskip 

Thus in integrating the dilation current over the boundary at 
infinity we are left with the term
\begin{equation}
S_{E}^{dil} = 
-\frac{\beta}{4\pi G_{d} \sqrt{d-2}} \int_{\partial\Sigma_{\infty}} 
(n \cdot \partial \phi) \sqrt{c}.
\end{equation}
Using the form of the metric, we can express this as 
\begin{eqnarray}
S_{E}^{dil} &=& - \frac{\beta}{16 \pi G_{d}} \int_{\partial \Sigma_{\infty}} 
\sqrt{\bar{g}} \lbrace \frac{1}{AB} \frac{\partial}{\partial r}(B^2)
- \frac{1}{A_{0}B_{0}} \frac{\partial}{\partial r}(B_{0}^2) 
\rbrace_{r=R}; \nonumber \\
&=& - \frac{\beta}{16 \pi G_{d}} \int_{\partial \Sigma_{\infty}} 
\sqrt{\bar{g}} R^{2n}
\lbrace \frac{\partial}{\partial r}(- \frac{\alpha}{r^{2n-1}} + O(\frac{1}
{r^{2n}})) \rbrace_{r=R}; \label{fzzn} \\
&= & \frac{\beta \alpha}{16 \pi G_{d}} \bar{V}_{2n} (2n-1); \nonumber \\
&=& \frac{\beta M}{2n} (2n-1). \nonumber
\end{eqnarray}
Adding the dilation current term to the surface gravity term, we find
that the action can be expressed as
\begin{equation}
S_{E} = \beta M - \frac{\beta}{8 \pi G_{d}} \int_{\partial \Sigma_{f}} 
J^{i}_{D} d\sigma_{i}. \label{defex}
\end{equation}
This implies that the entropy of the bolt solution with respect to the 
background nut solution is 
\begin{equation}
S = \sum_{a} \frac{V_{a}}{4 G_{d}} + \frac{\beta}{16 \pi G_{d}} \sum_{a}
\int_{M_{a}^{d-2}} F \wedge \Psi, \label{fixse}
\end{equation}
where we are implicitly subtracting the background quantities. 
Now using the simple cohomology structure of the base manifold to
evaluate the expression $(\ref{fixse})$ we find that 
\begin{equation}
S = \frac{(n(n+2)q^2)^n}{4G_{d}} \bar{V}_{2n} -
\frac{(n+1)q^3}{G_{d}} \bar{V}_{2n} [\Psi_{b} - \Psi_{n}],
\end{equation}
where $\Psi_{b,n}$ are the potentials evaluated at the fixed point
sets, such that 
\begin{equation}
\Psi_{b} - \Psi_{n} = [ \int^{q(n+1)}_{q}
\frac{p_{2n}(r)}{(r^2-q^2)^2} dr  + \frac{\alpha}{n(n+2)q^2}].
\end{equation}
It is non trivial to demonstrate that the total entropy is indeed 
given by (\ref{fzzn}); one needs to use the power series expansion for
the polynomial. 

\bigskip

One might wonder whether it were possible to match the nut
solutions to backgrounds which have no fixed point sets but the requisite
Dirac string behaviour. If one could, one could define the entropy and
mass of the nut solution with respect to this background, although the
background would not be a solution of the field equations. This is
certainly possible for the four dimensional Taub-Nut solution; the 
definition of the mass of the Kaluza-Klein monopole with respect to
such a background was discussed in \cite{BKKLS}. One takes the
background to be
\begin{equation}
ds^2_{4} = (d\tau + 2q \cos\theta d\psi)^2 + dr^2 + r^2
d\Omega_{2}^{2},
\end{equation}
which is the asymptotic form of the Taub-Nut metric. Although this
background is not flat, it is asymptotically locally flat in the sense
that the Ricci tensor falls off as $1/r^2$, and the volume term in the
action still vanishes. One can then define the
mass and entropy of Taub-Nut with respect to this background
as $q/G_{4}$ and $4\pi q^2/G_{4}$. However for $n>1$ solutions the
energies and actions defined with respect to such backgrounds are not
finite; the divergent terms do not cancel, and such backgrounds are
not appropriate.  

\section{Charge definition} \label{cgd}
\noindent

We will now consider the generalisation to gauge field theories. Let
us take a generic action of the form 
\begin{equation}
S_{E} = - \frac{1}{16\pi G_{d}} \int_{M} d^dx \sqrt{\hat{g}} [R - 
(\partial \Phi)^2 - e^{-a \Phi} {\mathcal H}_{p+1}^2], \label{dpact}
\end{equation}
where $\Phi$ is the $d$ dimensional dilaton, and ${\mathcal H}_{p+1}$
is a $(p+1)$ form, with the constant $a$ in general being 
dependent on $d$ and $p$. Using the equations of motion we can
express the volume term as 
\begin{equation}
S_{E} =  \frac{1}{8 \pi G_{d}} \int d^{d}x \sqrt{\hat{g}}
[\frac{p}{(d-2)} e^{-a \Phi} {\mathcal H}_{p+1}^{2} ].
\end{equation}
We will be interested in those solutions for which the metric is the radial
extension of a bundle over a homogeneous space. If the bundle is trivial,
we will obtain black hole solutions, whilst non-singular solutions with 
non-trivial bundle will be Israel-Wilson type metrics \cite{IsWi}.

In our previous paper \cite{MMT} we assumed that the $(d-1)$ dimensional gauge
field vanishes for compact manifolds with non-trivial gauge fields in $d$
dimensions. When we consider non compact solutions, we must however relax
this condition, since it is certainly possible to find a suitable Lorentzian
continuation on which all fields are real, by analytically continuing the mass,
nut parameter and gauge fields. Indeed the exclusion of $(d-1)$ dimensional
gauge fields on the Euclidean section corresponds to vanishing angular
momentum in the Lorentzian continuation. 
Thus it is important to consider the 
decomposition of the gauge field term for such solutions. 

\bigskip

We consider a metric of the form (\ref{line-el}), where we 
interpret the Killing vector as imaginary time.
Let the ``electric'' part of the $(p+1)$ form be 
${\mathcal H}_{(e)i_{1}..i_{p}} \equiv {\mathcal H}_{\tau
  i_{1}..i_{p}}$, and the ``magnetic'' part of the $(p+1)$ form be
${\mathcal H}_{(m)i_{1}..i_{p+1}} \equiv {\mathcal H}_{i_{1}..i_{p+1}}$.
So we can rewrite the integral of the $(p+1)$ form in terms of the
$(d-1)$ dimensional fields as 
\begin{eqnarray}
\int_{M} d^dx \sqrt{\hat{g}} e^{-a\Phi} {\mathcal H}_{p+1}^{2}
 \rightarrow  \beta \int_{\Sigma} d^{d-1}x \sqrt{g} e^{-a\Phi} 
\lbrace 
e^{-\frac{4 \phi (d+p-2)}{\sqrt{d-2}(d-3)}} (p+1) {\mathcal H}_{e}^{2}
\label {gyt} \\  
 + e^{-\frac{4 p \phi}{\sqrt{d-2}(d-3)}} \left [(p+1) {\mathcal H}_{e}^{2}A^2
+ ((p+1){\mathcal H}_{e} \cdot A)^{2} + {\mathcal H}_{m}^{2} +
(({\mathcal H}_{e} \cdot A) \cdot {\mathcal H}_{m}) \right]
\rbrace. \nonumber
\end{eqnarray}
This decomposition includes couplings between the Kaluza-Klein gauge
field and the $(d-1)$ dimensional $p$ form and $(p+1)$ form, but all
but one term can be replaced by the dilation current, since its
divergence is 
\begin{eqnarray}
D_{i}J^{i}_{D} = - \frac{p}{d-2} e^{-a \Phi} 
e^{-\frac{4 p \phi}{\sqrt{d-2}(d-3)}} [(p+1) {\mathcal H}_{e}^{2}A^2 +
....] \\
- \frac{(d+p-2)}{d-2} (p+1) e^{-a \Phi} 
e^{-\frac{4 \phi (d+p-2)}{\sqrt{d-2}(d-3)}} {\mathcal H}_{e}^{2},
\nonumber 
\end{eqnarray}
where the dots indicate the remainder of the terms contained in
square brackets in (\ref{gyt}).
Thus we can write the volume term in the on shell action as 
\begin{equation}
S_{E} = - \frac{\beta}{8 \pi G_{d}} \int_{\Sigma} D_{i} J^{i}_{D} 
+ \frac{1}{8 \pi G_{d}} \int_{M} d^{d}x \sqrt{\hat{g}} e^{-a\Phi}
((p+1) g_{\tau\tau}^{-1} {\mathcal H}_{e}^{2}),
\end{equation}
where in the latter expression we define ${\mathcal H}_{e}^{2}$ in the
original $d$ dimensional metric. Now, for the class of metrics that we
are considering, black hole solutions and generalised Taub-Nut
solutions, the dilation current and surface gravity terms reduce
to terms involving the mass and sums over fixed point sets as
previously. That is, 
\begin{equation}
S_{E} = \beta M - S_{f} 
+ \frac{1}{8 \pi G_{d}} \int_{M} d^{d}x \sqrt{\hat{g}} e^{-a\Phi}
((p+1) g_{\tau\tau}^{-1} {\mathcal H}_{e}^{2}), \label{decz}
\end{equation} 
where $S_{f}$ is the usual sum over fixed point sets. 

\bigskip

Now (\ref{decz}) certainly implies
the well known results for electric and magnetic
dilatonic black holes. If the gauge field is pure magnetic, then the 
action can be expressed as 
\begin{equation}
S_{E} = \beta M - \frac{A_{2}}{4 G_{4}},
\end{equation}
where $A_{2}$ is the area of the horizon and
we restrict to four dimensions since magnetic charge is not
defined in higher dimensions. If the gauge field is pure electric, and
the bundle over the base manifold is trivial, then the integral in
(\ref{decz}) reduces to 
\begin{equation}
\frac{1}{8 \pi G_{d}} \int_{M} d^{d}x \sqrt{\hat{g}} e^{-a\Phi} \bar{F}^{2} = 
\frac{1}{4\pi G_{d}} \int_{\partial M} d^{d-1}x \sqrt{b}
e^{-a\Phi} \bar{F}^{\mu\nu} n_{\mu} \bar{A}_{\nu}, 
\end{equation}
where $n_{\mu}$ is the normal to the boundary. We can 
use the field equations to convert the volume term to a
surface term provided that there are no Dirac string 
singularities. An appropriate choice
of gauge is such that $\bar{A}_{\mu}$ vanishes on the inner boundary, and
relating the integral over the boundary to the charge we find that 
\begin{equation}
S_{E} = \beta M - \beta q \chi- \frac{V_{h}}{4G_{d}},
\end{equation}
where $\chi$ is the gauge potential at infinity, $q$ is the
(suitably normalised) charge and
$V_{h}$ is the volume of the horizon. As usual,
we can introduce a boundary term in the action for an electric
solution \cite{HR}, so
that the actions for magnetic and electric black holes are equal for
equal charge in four dimensions. Note that although the differential
form of the first law of black hole dynamics does depend not only on
the charges, but also on the scalar fields, the integrated version is 
independent of these fields, and we would not expect to see any
such contributions appearing in the action. 

\bigskip

We should note that for a black $(p-1)$-brane solution of topology
$R^{p+1} \times S^{d-p-1}$ carrying electric charge with respect to
the $(p+1)$ form the integral over
the gauge field reduces to 
\begin{equation}
\frac{1}{8 \pi G_{d}} \int_{M} d^dx \sqrt{\hat{g}} e^{-a \Phi}
{\mathcal H}_{p+1}^{2} = \frac{(p+1)}{4\pi G_{d}} \int_{\partial M} (\ast
e^{-a\Phi} {\mathcal H}) \wedge {\mathcal B}_{p}
= - \beta (\prod_{i=1}^{p-1} \beta_{i}) B_{\infty} \varrho,
\end{equation}
where the charge per unit area of the $(p-1)$-brane is defined
according to the convention 
\begin{equation}
\varrho = \frac{(p+1)}{8\pi G_{d}} \int_{S^{d-p-1}} (\ast e^{-a\Phi}
{\mathcal H}),
\end{equation}
and $\beta_{i}$ are the dimensions of the $(p-1)$-brane. The gauge
potential is chosen to vanish on the horizon, and takes the form
\begin{equation}
{\mathcal B}_{p} = B_{\infty} dt \wedge \prod_{i=1}^{p-1} dx^{i},
\end{equation}
at the boundary at
infinity, where the $x^{i}$ are longitudinal coordinates on the
brane. Thence 
\begin{equation}
S_{E} = \beta M - S_{f} - \beta Q B_{\infty},
\end{equation}
where $Q$ is the total charge of the $(p-1)$-brane, and $M$ is the
total mass. Then using the explicit expression for the action we can
show that
\begin{equation}
(\frac{d-3}{d-2}) \beta M = S_{f} + (\frac{d-2-p}{d-2}) \beta Q
B_{\infty},
\end{equation}
which is the generalisation \cite{GibbMae} 
of the Smarr law for static electric black branes, as required.  

\bigskip

If the bundle over the homogenous manifold is trivial, the two form 
term can be expressed in terms of the charge and horizon potential. If however
the $(d-1)$ dimensional gauge field is non trivial, the charge is not
well defined. That is, one usually requires a $(d-2)$ dimensional sphere at 
infinity over which we integrate the dual of the gauge field in order to define
electric charge. Even if we consider radially extended bundles over other types
of homogeneous manifolds, such as projective spaces, one can define a charge
by integrating the dual field over a $(d-2)$ dimensional boundary at infinity,
unless the bundle over the homogenous manifold is non trivial, in which case 
there will be no topologically suitable surfaces over which we can integrate.

Thus we must replace the charge and potential term which can only be
defined for radially extended trivial bundles
with an appropriate volume integral of the gauge field. 
Although the surface integral used to evaluate the electric charge is
not well defined since there do not extend topologically suitable
boundaries over which one can integrate, the volume integral is well
defined. 

\bigskip

Now for solutions of the equations of motion derived from the action
(\ref{dpact}) we can derive the equivalent of the Smarr law for 
black holes. Assuming that the solution is static, 
\begin{eqnarray}
(\frac{d-3}{d-2}) \beta M = S_{f} + 
\frac{1}{8 \pi G_{d}} \int_{M} d^dx \sqrt{\hat{g}} (\frac{p}{d-2}) e^{-a \Phi}
{\mathcal H}_{p+1}^{2} \\
- \frac{1}{8\pi G_{d}} \int_{M} d^dx
\sqrt{\hat{g}} (p+1) e^{-a\Phi} g_{\tau\tau}^{-1} {\mathcal
  H}_{e}^{2}. \nonumber
\end{eqnarray}
We can thence use the well
defined volume integral to extend the decomposition of the action to
non trivial topologies in which one cannot define electric (or
magnetic) charges. Note that although one can formally express the
volume term for the solution in this way one cannot guarantee that
each term is finite, but the difference between the solution and
background contributions should be finite if there is exist a sensible
thermodynamic interpretation. 

\section{Dyonic Taub-Nut solutions} \label{dtn}
\noindent

To illustrate the results of the previous section, 
we are interested in solutions which are not asymptotically flat, for
which the gauge fields are non trivial but the charge is not well
defined. We could also use the Israel Wilson family of solutions to
discuss such behaviour, but will instead
consider the Euclidean sections of Taub-Nut
dyon solutions constructed as solutions to an effective action of four 
dimensional heterotic string theory. The (truncated) action takes the form
\begin{equation}
S_{E} = - \frac{1}{16 \pi G_{4}} \int_{M} d^4 x \sqrt{\hat{g}} [R - 
\frac{1}{2} (\partial \Phi)^2 - \frac{1}{8} e^{-\Phi} \bar{F}^2 - \frac{1}{12}
e^{-2\Phi} H^2],
\end{equation}
where as usual $\Phi$ is the dilaton, $\bar{F}$ is a two form and $H$ is
a three form field. Enforcing the on shell conditions and including the surface
term one finds that the action is
\begin{equation}
S_{E} = \frac{1}{16 \pi G_{4}} \int_{M} d^4 x \sqrt{\hat{g}} 
(\frac{1}{8} e^{-\Phi} \bar{F}^2 + \frac{1}{6} e^{-2\Phi} H^2) - 
\frac{1}{8\pi G_{4}} \int_{\partial \Sigma} 
({\mathcal K} - {\mathcal K}_{0}) \sqrt{c}.
\end{equation}
Note that the Chern-Simons term in the definition of the three form 
$H = dB + \omega(\bar{A})$, where 
\begin{equation}
\omega_{\mu\nu\rho} = \frac{1}{4} (\bar{A}_{\mu} \bar{F}_{\nu\rho} +
\bar{A}_{\nu}\bar{F}_{\rho\mu} + \bar{A}_{\rho}\bar{F}_{\mu\nu}),
\end{equation}
will not affect our decomposition of the action, since at no stage did
we assume either that $H$ is locally exact or that $d\ast H = 0 =
d\ast \bar{F}$. 
The class of solutions that we consider was constructed in \cite{JMy} 
from $O(1,1)$ T-duality transformations on the
Lorentzian Taub-Nut solutions. The Lorentzian fields are 
\begin{eqnarray}
ds^2 &=& - \frac{f_{1}}{f_{2}} (dt + q (x+1) \cos\theta d\psi)^2 + 
\frac{f_{2}}{f_{1}} dr^2 + f_{2} (r^2 + q^2) d\Omega_{2}^{2}; \nonumber\\
\bar{A}_{t} &=&  \sqrt{x^2 - 1} \frac{(1 - f_{1})}{f_{2}}; \nonumber \\
\bar{A}_{\psi} &=& - 2 \frac{f_{1}}{f_{2}} \sqrt{x^2-1} q \cos\theta; \\
B_{t\psi} &=& \frac{f_{1}}{f_{2}} (x-1)  q \cos\theta; \nonumber \\
e^{-\Phi} &=& f_{2} \nonumber,
\end{eqnarray}
where we give the Einstein frame metric. The functions are defined as 
\begin{eqnarray}
f_{1}  &=& 1 - 2 \frac{Mr + q^2}{(r^2 + q^2)}; \label {futy} \\
f_{2} &=& 1 + (x-1) \frac{Mr + q^2}{(r^2 + q^2)}. \nonumber
\end{eqnarray}
The parameter $x$ is a boost such that 
$x^2 \ge 1$, and the time coordinate must be identified with a period 
$4 \pi q (x+1) $ to avoid Dirac string singularities. 
To find the Euclidean section of the original family
of solutions, we must let the nut parameter $q$ become imaginary
whilst letting the mass $M$ remain real. Taking the same approach here we
find that
\begin{eqnarray}
ds^2 &=& \frac{F_{1}}{F_{2}} (d\tau + q (x+1) \cos\theta d\psi)^2 + 
\frac{F_{2}}{F_{1}} dr^2 + F_{2} (r^2 - q^2) d\Omega_{2}^{2};
\nonumber \\
\bar{A}_{\tau} &=& i \bar{A}_{t}; \ \bar{A}_{\psi} \rightarrow i
\bar{A}_{\psi}; \ B_{\tau\psi} = - B_{t\psi}
\end{eqnarray}
where the functions $F_{1}$, $F_{2}$ are found by continuing the
parameters of (\ref{futy}). Requiring the Dirac
string singularity to be removable implies that $\tau$ must be identified
with a periodicity $\beta = 4 \pi q (x+1)$ in both the Euclidean and the 
Lorentzian solutions. Regularity at the origins, the fixed point sets,
then requires that $M$ must take the values $q$ or $5q/4$ in the Euclidean
solution, as in the usual Taub-Nut and Taub-Bolt solutions. The fixed
points of both regular solutions exhibit the same behaviour as in the
usual solutions. 

There will also be an appropriate limit in which the nut parameter
$q$ vanishes but the mass does not; one then obtains black hole
solutions, and can take bundle over the base manifold to be trivial. 
These are precisely the family of extreme solutions considered 
in \cite{GibbMae}. Note that the 
boosted solutions do not permit interpretations as monopoles;
the Lorentzian solutions obtained by the addition of a flat time
direction are complex. 

Usually when one has an electrically charged solution the electric
part of the field becomes imaginary in the Lorentzian continuation,
and the magnetic part of the field remains real. Here we find that
the one form potential is pure imaginary but the two form is pure real
real. This is because we have let not only the time coordinate, but
also the nut parameter, become pure imaginary in the Euclidean
continuation.

We should also mention that the Euclidean solution for which $x= - 1$
is singled out. For this particular choice, the non trivial fibration
of the imaginary time coordinate over the spatial two sphere is lost,
and the topology of surfaces of constant $r$ is simply $S^1 \times
S^2$. One obtains this effect by
allowing for discrete duality transformations; however, one then
finds that there exist curvature singularities in the Lorentzian solutions 
which are not concealed by horizons. 

\bigskip

The simplest way to calculate the action is to use the string frame
metric, since one can then use the dilaton equation of motion
to convert the volume integral to a surface term and thence
\begin{equation}
S_{E} = - \frac{1}{8 \pi G_{4}} \int_{\partial M} [n \cdot 
\partial (e^{-\Phi}) +
({\mathcal{K}} - {\mathcal {K}}_{0})] \sqrt{C} d^3 x,
\end{equation}
where we implicitly subtract the background dilaton divergence 
term, and $C$ is the induced metric on the boundary in the string
frame. Using the usual approach of 
matching of the bolt metric to a background nut
metric, we can evaluate the action to be
\begin{equation}
S_{E} = \frac{\beta q}{8 G_{4}} = \frac{\pi q^2 (1+x)}{4G_{4}},
\end{equation}
which is in agreement with that of the $x=1$ uncharged solutions. 
Note that the
total action decreases as we increase the boosting of the solutions,
and thus the formal probability for decay decreases (although the
decay is still prohibited by cobordism arguments).  

Working in the Einstein frame, we can determine 
the mass of the bolt solution with
respect to the nut solution as $q(1+x)/8 G_{4}$ so that
\begin{equation}
\beta E = \frac{\pi q^2 (1+x)^2}{4 G_{4}},
\end{equation}
and the fixed point set term can be evaluated to be
\begin{equation}
S_{f} = \frac{3\pi q^2}{2 G_{4}} (1+x) - \frac{\pi q^2}{G_{4}}
 (1+x)^2 [1 - \Psi_{B}],
\end{equation}
where $\Psi_{B}$ is the potential evaluated at the bolt
\begin{eqnarray}
\Psi_{B} &=& \frac{4(x+1)}{f^2} \lbrace \frac{36}{f} \tanh^{-1}
(\frac{f}{5x+11}) (1-x) + (5x-1) \rbrace; \\
f &=& (25 + 14x + 25 x^2)^{1/2}.
\end{eqnarray}
Since $\Psi_{B} \rightarrow 0.8$ for large $x$, we can see immediately
that the entropy of the bolt solution with respective to the nut
solution will be negative for large $x$, and the bolt solution will be
much heavier. Hence the formal decay rate must
be heavily suppressed as indeed we found from the action. 

\section{Conclusions} \label{fin}
\noindent

In this paper, we have discussed the action of a circle isometry group
on non-compact Euclidean Einstein manifolds and Euclidean
solutions of supergravity theories. The analysis is more subtle than
that for compact manifolds discussed in \cite{MMT} and in \cite{GH}. 
Firstly, we require the existence of a suitable background with 
respect to which all thermodynamic quantities can be defined. Secondly,
we have to identify an appropriate temperature before we can define
the entropy; we must work within a canonical or grand canonical
ensemble. When we attempt to decompose the action in terms of the
fixed point sets of the action of an isometry in general there will
be surface terms on the boundary at infinity which are left
over. 

We can define a decomposition of the action in terms of the fixed point
sets of any Killing vector but for this decomposition to have any
physical significance these fixed point sets must be null in the
Lorentzian continuation. A good example is a Kerr solution; one could
define the decomposition of the action in terms of the fixed point
sets of the imaginary time Killing vector, but it is the fixed point
sets of the generator of the event horizon which are null in the
Lorentzian continuation, and are associated with an entropy and a
temperature. 

Now if one does express the action of the solution in terms of the
fixed point sets of an appropriate Killing vector which admits null
horizons one might expect that the surface terms at infinity are
related to the mass and angular momentum of the solution with the
fixed point set terms relating to the entropy. That is, the action
would take the form
\begin{equation}
S_{E} = \beta E - \beta \omega_{i} J_{i} - S_{f}, \label{ppy}
\end{equation}
where the quantity $S_{f}$ will be interpreted as an entropy and is
defined as 
\begin{equation}
S_{f} = \sum_{a} \frac{V_{a}}{4G_{d}} + \frac{\beta}{16 \pi G_{d}}
\sum_{a} \int_{M_{a}^{d-2}} F \wedge \Psi. \label{stet}
\end{equation}
This expression is of course well known for general black brane
solutions, for which the bundle over the base manifold is trivial, and
thence the integral of the potential vanishes. It would be easy to
demonstrate within the formalism used here that this expression for
the action of generic rotating black branes holds by assuming
an appropriate general form for the metric. 

\bigskip

To demonstrate that this
expression holds in general is non trivial without imposing an ansatz
for solution and background; indeed we would expect this, since for
solutions which are not asymptotically flat the definition of the
energy is highly non trivial. Thus we restrict to the most simple case,
that of static solutions admitting a single fixed point set with Dirac
string type behaviour. To define thermodynamic quantities one must
compare a solution with a background which has identical magnetic
behaviour, but which is in some sense more symmetric. 

In four dimensions, the appropriate Euclidean solutions are the
Taub-Bolt and Taub-Nut manifolds, for which the Lorentzian continuations are
members of the Taub-Nut family with particular masses. In higher
dimensions, the Euclidean section of the generalised Taub-Nut solution
had been constructed, but the corresponding generalised Taub-Bolt
solution was not known to exist. Here we have constructed the
higher dimensional generalisation of Taub-Bolt, and calculated the
action of the latter with respect to a nut background. We also
considered the Lorentzian continuation of such solutions, which
exhibit very similar behaviour to the four dimensional family of
solutions. 

Given such solutions and appropriate backgrounds, we demonstrate
explicitly that the surface integrals at infinity reduce to the mass 
which implies that the entropy takes the form (\ref{stet}). Since the 
two form and potential are non trivial, we have a generalised nut
type contribution to the entropy, as well as the contribution from the
horizon volume. Now in principle we could consider rotating these
nut solutions and we would expect that infinitesimal rotations would
preserve the form of the entropy. So for more general solutions we
would expect the expression (\ref{ppy}) to hold. 

Having considered a classification scheme of non compact Euclidean Einstein
manifolds it is natural to again extend the ideas to solutions of
gravity coupled to appropriate scalar and gauge fields. Again we find
that the decomposition of the action is unaffected, except for
electric fields when we will obtain an additional integral left
over. For black brane solutions, this term is related to the charge
and an appropriate potential; by introducing a surface term at
infinity, one can remove this term in the action. The physical
interpretation of removing this term 
is that one then considers only variations of the gauge
field which leave the electric charge unchanged. 

\bigskip

For more general solutions, which display a non-trivial Dirac string 
behaviour, the additional term in the action cannot be related to a
surface integral over the boundary. Furthermore, one cannot define
either an electric or a magnetic charge, since there will exist no
topologically non trivial surfaces at infinity 
over which we can integrate the appropriate forms. So for such
manifolds we cannot simply add a boundary term relating to a charge
and a potential to cancel that arising from the decomposition of the
volume term. Nevertheless the entropy is well defined if we interpret
this additional volume term as a constraint. 

As an example of such behaviour, we considered exact solutions of
four dimensional heterotic string theory, obtained from the Taub-Nut
family of solutions by applying T-duality transformations. Such 
solutions have non trivial dyonic two forms, and an electric three
form. Charge cannot be defined, but we were still able to define the
entropy of such a solution with respect to an appropriate background. 

One might claim that since any solution which has a non zero
contribution to the entropy from Dirac string behaviour is by
definition not asymptotically flat, or even locally flat, except in
four dimensions, that such solutions have little physical relevance. 
In some sense, perhaps one should only consider their creation in
pairs, for which there will be no net Dirac string. However, such
solutions are certainly appropriate backgrounds in string theory, even
those obtained by T-duality transformations, and should perhaps not be
neglected.

\end{document}